\documentclass[aip,apl,preprint]{revtex4-1}
\usepackage{graphics}
\usepackage{mathtools}

\begin{document}
\title{High-efficiency GHz frequency doubling without power threshold in thin-film Ni$_{81}$Fe$_{19}$}
\author{Cheng Cheng}
\author{William E. Bailey}
\affiliation{Materials Science and Engineering Program, Department of Applied Physics and Applied Mathematics, Columbia University, New York, NY 10027}

\begin{abstract} 
We demonstrate efficient second-harmonic generation at moderate input power for thin film Ni$_{81}$Fe$_{19}$ undergoing ferromagnetic resonance (FMR). Powers of the generated second-harmonic are shown to be quadratic in input power, with an upconversion ratio three orders of magnitude higher than that demonstrated in ferrites \cite{AyresJAP1956}, defined as $\Delta P^{2\omega}/\Delta P^{\omega} \sim 4 \times 10^{-5} /W \cdot P^{\omega}$, where $\Delta P$ is the change in the transmitted rf power and $P$ is the input rf power. The second harmonic signal generated exhibits a significantly lower linewidth than that predicted by low-power Gilbert damping, and is excited without threshold. Results are in good agreement with an analytic, approximate expansion of the Landau-Lifshitz-Gilbert (LLG) equation.
\end{abstract}

\maketitle

Nonlinear effects in magnetization dynamics, apart from being of fundamental interest\cite{WangPR1954, AyresJAP1956, SuhlJPCS1957, BierleinPRB1970}, have provided important tools for microwave signal processing, especially in terms of frequency doubling and mixing\cite{RidrigueJAP1969, HarrisIEEE2012}. Extensive experimental work exists on ferrites\cite{AyresJAP1956, BierleinPRB1970, HarrisIEEE2012}, traditionally used in low-loss devices due to their insulating nature and narrow ferromagnetic resonance (FMR) linewidth. Metallic thin-film ferromagnets are of interest for use in these and related devices due to their high moments, integrability with CMOS processes, and potential for enhanced functionality from spin transport; low FMR linewidth has been demonstrated recently in metals through compensation by the spin Hall effect\cite{DemidovNMat2012}. While some recent work has addressed nonlinear effects\cite{BerteaudJAP1966, GerritsPRL2007, OlsonJAP2007} and harmonic generation\cite{BaoAPL2008, KhivintsevAPL2011, MarshAPL2012} in metallic ferromagnets and related devices\cite{YanaJMMM2008,DemidovAPL2011, BiAPL2011}, these studies have generally used very high power or rf fields, and have not distinguished between effects above and below the Suhl instability threshold. In this manuscript, we demonstrate frequency doubling below threshold in a metallic system (Ni$_{81}$Fe$_{19}$) which is three orders of magnitude more efficient than that demonstrated previously in ferrite materials\cite{AyresJAP1956}. The results are in good quantitative agreement with an analytical expansion of the Landau-Lifshitz-Gilbert (LLG) equation. \\

For all measurements shown, we used a metallic ferromagnetic thin film structure, Ta(5 nm)/Cu(5 nm)/Ni$_{81}$Fe$_{19}$
(30 nm)/Cu(3 nm)/Al(3 nm). The film was deposited on an oxidized silicon substrate using magnetron sputtering at a base pressure of 2.0$\times$10$^{-7}$ Torr. The bottom Ta(5 nm)/Cu(5 nm) layer is a seed layer to improve adhesion and homogeneity of the film and the top Cu(3 nm)/Al(3 nm) layer protects the Ni$_{81}$Fe$_{19}$ layer from oxidation. A diagram of the measurement configuration, adapted from a basic broadband FMR setup, is shown in Fig.1. The microwave signal is conveyed to and from the sample through a coplanar waveguide (CPW) with a 400 $\mu$m wide center conductor and 50 $\Omega$ characteristic impedance, which gives an estimated rf field of 2.25 Oe rms with the input power of +30 dBm. We examined the second harmonic generation with fundamental frequencies at 6.1 GHz and 2.0 GHz. The cw signal from the rf source is first amplified by a solid state amplifier, then the signal power is tuned to the desirable level by an adjustable attenuator. Harmonics of the designated input frequency are attenuated by the bandpass filter to less than the noise floor of the spectrum analyzer (SA). The isolator limits back-reflection of the filtered signal from the sample into the rf source. From our analysis detailed in a later section of this manuscript, we found the second harmonic magnitude to be proportional to the product of the longitudinal and transverse rf field strengths, and thus place the center conductor of CPW at 45$^{\circ}$ from H$_B$ to maximize the $H^{rf}_y H^{rf}_z$ product. The rf signal finally reaches the SA for measurements of the power of both the fundamental frequency and its second harmonic.\\
\begin{figure}
	\includegraphics{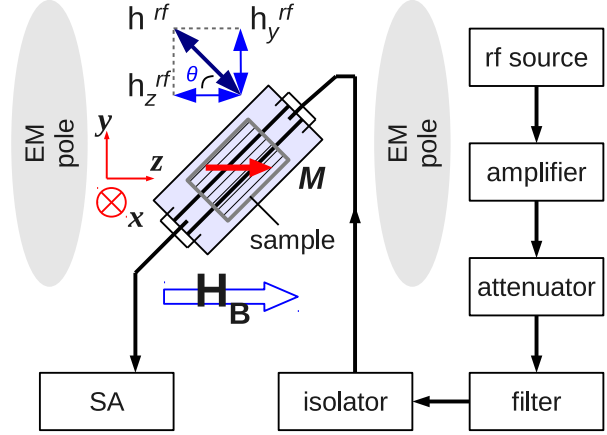}
	\caption{\label{Fig.1} Experimental setup and the coordinate system, $\theta = 45^{\circ}$; see text for details. EM: electromagnet; SA: spectrum analyzer. Arrows indicate the transmission of rf signal.}
\end{figure}

\begin{figure}
	\includegraphics{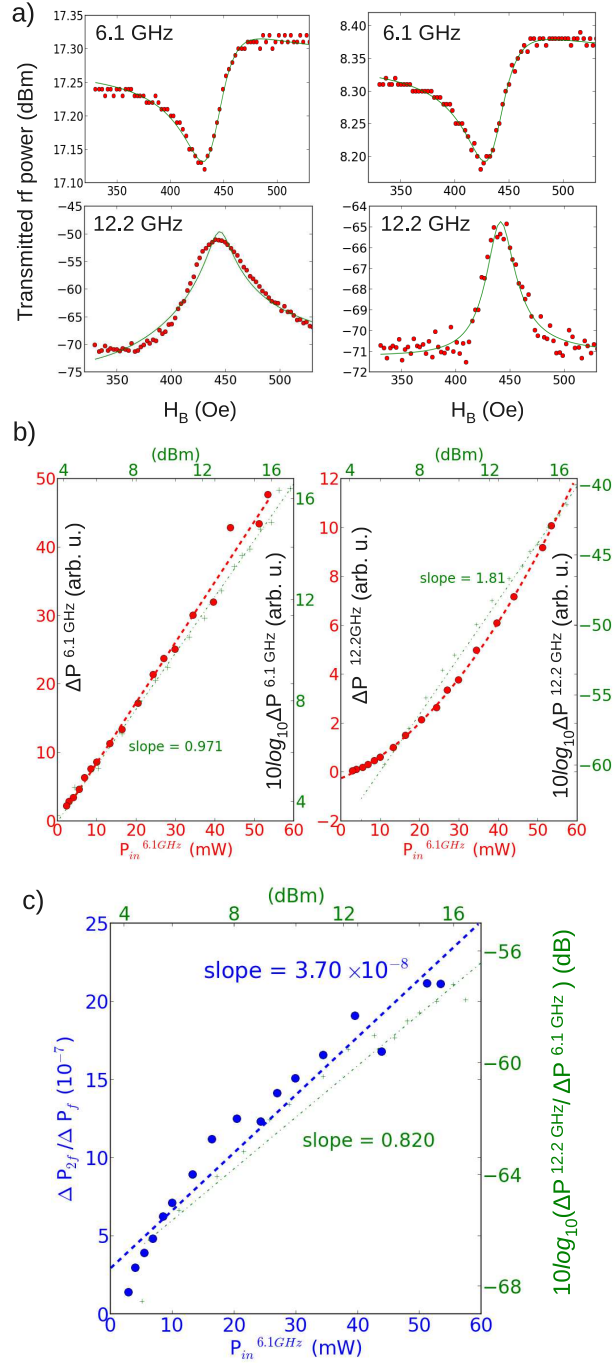}
	\caption{\label{Fig.2} Second harmonic generation with $\omega/2\pi = $ 6.1 GHz. a) \textit{left panel}: 6.1 GHz input power +17.3 dBm; \textit{right panel}: 6.1 GHz input power +8.35 dBm. b) amplitudes of the $\omega$ (FMR) and generated 2$\omega$ peaks as a function of input power $P^{\omega}$; right and top axes represent the data set in log-log plot (green), extracting the power index; c) ratio of the peak amplitudes of FMR and second harmonic generation as a function of the input 6.1 GHz power; green: log scale.}
\end{figure}
\indent Fig.2(a) demonstrates representative field-swept FMR absorption and the second harmonic emission spectra measured by the SA as 6.1 GHz and 12.2 GHz peak intensities as a function of the bias field H$_B$. We vary the input rf power over a moderate range of +4 - +18 dBm, and fit the peaks with a Lorentzian function to extract the amplitude and the linewidth of the absorbed ($\Delta P^{\omega}$) and generated ($\Delta P^{2\omega}$) power. Noticeably, the second harmonic emission peaks have a much smaller linewidth, $\Delta H_{1/2}\sim$ 10 Oe over the whole power range, than those of the FMR peaks, with $\Delta H_{1/2}\sim$ 21 Oe. Plots of the absorption and emission peak amplitudes as a function of the input 6.1 GHz power, shown in Fig.2(b), clearly indicate a linear dependence of the FMR absorption and a quadratic dependence of the second harmonic generation on the input rf power. Taking the ratio of the radiated second harmonic power to the absorbed power, we have a conversion rate of 3.7$\times$10$^{-5}$/W, as shown in Fig.2(c). \\
\indent Since the phenomenon summarized in Fig.2 is clearly not a threshold effect, we look into the second-harmonic analysis of the LLG equation with small rf fields, which is readily described in Gurevich and Melkov's text for circular precession relevant in the past for low-M$_s$ ferrites\cite{Gurevich&Melkov}. For metallic thin films, we treat the elliptical case as follows. As illustrated in Fig.1, the thin film is magnetized in the \textit{yz} plane along $\widehat{\textbf{z}}$ by the bias field H$_B$, with film-normal direction along $\widehat{\textbf{x}}$. The CPW exerts both a longitudinal rf field h$_z^{rf}$ and a transverse rf field h$_y^{rf}$ of equal strength. First consider only the transverse field h$_y^{rf}$. In this well established case\cite{}, the LLG equation $\dot{\textbf{m}} = -\gamma \textbf{m}\times \textbf{H}_{eff} + \alpha \textbf{m}\times \dot{\textbf{m}}$
is linearized and takes the form
\begin{equation}
  \begin{bmatrix}
    \dot{\widetilde{m_x}}   \\
    \dot{\widetilde{m_y}}
  \end{bmatrix}
  =\begin{bmatrix}
    -\alpha(\omega_H+\omega_M) & -\omega_H  \\
    \omega_H+\omega_M & -\alpha\omega_H
  \end{bmatrix}
  \begin{bmatrix}
    \widetilde{m_x}  \\
    \widetilde{m_y}
  \end{bmatrix}
  +\begin{bmatrix}
    \gamma \widetilde{h_y^{rf}} \\
    0
  \end{bmatrix}
\end{equation}
, where $\gamma$ is the gyromagnetic ratio, $\alpha$ is the Gilbert damping parameter, $\omega_M \equiv \gamma 4\pi M_s$, and $\omega_H \equiv \gamma H_z$. Introducing first order perturbation to $m_{x,y}$ under additional longitudinal h$_z^{rf}$ and neglecting the second order terms, we have 
\begin{equation}
  \begin{bmatrix}
    \dot{\widetilde{m_x}} +\dot{\widetilde{\Delta m_x}}\\
    \dot{\widetilde{m_y}} +\dot{\widetilde{\Delta m_y}}
  \end{bmatrix}
  =\begin{bmatrix}
    -\alpha(\omega_H+\omega_M) & -\omega_H  \\
    \omega_H+\omega_M & -\alpha\omega_H
  \end{bmatrix}
  \begin{bmatrix}
    \widetilde{m_x} + \widetilde{\Delta m_x} \\
    \widetilde{m_y} + \widetilde{\Delta m_y}
  \end{bmatrix}
  +\begin{bmatrix}
    \gamma \widetilde{h_z^{rf}} \widetilde{m_y}\\
    -\gamma \widetilde{h_z^{rf}} \widetilde{m_x}
  \end{bmatrix}
  +\begin{bmatrix}
    \gamma \widetilde{h_y^{rf}} \\
    0
  \end{bmatrix}
\end{equation}
Subtracting (1) from (2) and taking $\widetilde{h^{rf}}_{y,z} = H_{y,z}^{rf}e^{-i\omega t}$, $\widetilde{m}_{x,y} = (H_y^{rf}/M_s)e^{-i\omega t}\widetilde{\chi}_{\perp,\parallel}(\omega)$, the equation for the perturbation terms is 
\begin{equation}
  \begin{bmatrix}
    \dot{\widetilde{\Delta m_x}}\\
    \dot{\widetilde{\Delta m_y}}
  \end{bmatrix}
  =\begin{bmatrix}
    -\alpha(\omega_H+\omega_M) & -\omega_H  \\
    \omega_H+\omega_M & -\alpha\omega_H
  \end{bmatrix}
  \begin{bmatrix}
    \widetilde{\Delta m_x} \\
    \widetilde{\Delta m_y}
  \end{bmatrix}
  +H_z^{rf}\frac{H_y^{rf}}{M_s}e^{-i2\omega t}\begin{bmatrix}
    \gamma \widetilde{\chi_{\parallel}}(\omega)\\
    -\gamma \widetilde{\chi_{\perp}}(\omega)
  \end{bmatrix}
\end{equation}
Since $\chi_{\perp}$ is one order of magnitude smaller than $\chi_{\parallel}$, we neglect the term $-\gamma \widetilde{\chi_{\perp}}(\omega)$. In complete analogy to equation (1), the driving term could be viewed as an effective transverse field of $H_z^{rf}(H_y^{rf}/M_s)\widetilde{\chi_{\parallel}}(\omega)e^{-i2\omega t}$, and the solutions to equation (3) would be
$
  \widetilde{\Delta m_x} = (H_z^{rf}H_y^{rf}/M_s^2)\widetilde{\chi_{\parallel}}(\omega)\widetilde{\chi_{\perp}}(2\omega)e^{-i2\omega t} $, $
  \widetilde{\Delta m_y} = (H_z^{rf}H_y^{rf}/M_s^2)\widetilde{\chi_{\parallel}}(\omega)\widetilde{\chi_{\parallel}}(2\omega)e^{-i2\omega t}
$.
We can compare the power at frequency $f$ and $2f$ now that we have the expressions for both the fundamental and second harmonic components of the precessing \textbf{M}. The time-averaged power per unit volume could be calculated as 
$
  \langle P\rangle  = [\int_{0}^{\frac{2\pi}{\omega}} P(t) dt]/(2\pi/\omega), \ \ P(t) = -\partial U/\partial t = 2\textbf{M}\partial \textbf{H}/\partial t
$
where only the transverse components of \textbf{M} and \textbf{H} contribute to P(t). Using the expression for $ \langle P\rangle$, \textbf{M} and \textbf{H},  we have 
$
  P^{\omega}  = \omega H_{y,rf}^2\chi(\omega)''_{\parallel}
$
and
$
  P^{2\omega}  = 2\omega H_{z,rf}^2(H_y^{rf}/M_s)^2|\tilde{\chi}(\omega)_{\parallel}|^2\chi(2\omega)''_{\parallel}
$,
from which we conclude that under H$_B$ for FMR at frequency $f=\omega/(2\pi)$, we should see a power ratio 
\begin{equation}
P^{2\omega}/P^{\omega} = 2(H_z^{rf}/M_s)^2\chi(\omega)''_{\parallel}\chi(2\omega)''_{\parallel}
\end{equation}
With $M_s =$ 844 Oe, $\alpha =$ 0.007 as measured by FMR for our Ni$_{81}$Fe$_{19}$ 30 nm sample and 2.25 Oe rf field amplitude at input power of 1 W for the CPW, we have a calculated $2f/f$ power ratio of 1.72$\times$10$^{-5}$/W, which is in reasonable agreement with the experimental data 3.70$\times$10$^{-5}$/W as shown in Fig.2(c). To compare this result with the ferrite experiment in ref.[1], we further add the factor representing the ratio of FMR absorption to the input rf power, which is $3.9 \times 10^{-2}$ in our setup. This leads to an experimental upconversion ratio of $1.44 \times 10^{-6}$/W  in ref.[1]'s definition ($\Delta P^{2\omega} / {P^{\omega}_{in}}^2 $), compared with $7.1 \times 10^{-10}$/W observed in Mg$_{70}$Mn$_8$Fe$_{22}$O (Ferramic R-1 ferrite). \\
\indent Examining Eq.(4), we notice that there should be two peaks in the field-swept $2f$ emission spectrum: the first coincides with the FMR but with a narrower linewidth due to the term $|\tilde{\chi}(\omega)_{\parallel}|^2$, and the second positioned at the H$_B$ for the FMR with a $2f$ input signal due to the term $\chi(2\omega)''_{\parallel}$. The second peak should have a much smaller amplitude. Due to the field limit of our electromagnet, we could not reach the bias field required for FMR at 12.2 GHz under this particular configuration and continued to verify Eq.(4) at a lower frequency of 2.0 GHz. We carried out an identical experiment and analysis and observed an upconversion efficiency of 0.39$\times$10$^{-3}$/W for the 4.0 GHz signal generation at 2.0 GHz input, again in reasonable agreement with the theoretical prediction 1.17$\times$10$^{-3}$/W. Fig.3 demonstrates the typical line shape of the 4 GHz spectrum, in which the input 2 GHz power being +18.9 dBm. A second peak at the H$_B$ for 4 GHz FMR is clearly visible with a much smaller amplitude and larger linewidth than the first peak, qualitatively consistent with Eq.(4). A theoretical line (dashed green) from equation (4) with fixed damping parameter $\alpha = 0.007$ is drawn to compare with the experimental data. The observed second peak at the $2f$ resonance H$_B$ shows a much lower amplitude than expected. We contribute this difference to the possible $2f$ component in the rf source which causes the $2f$ FMR absorption. The blue line shows the adjusted theoretical line with consideration of this input signal impurity.\\ 
\begin{figure}
	\includegraphics{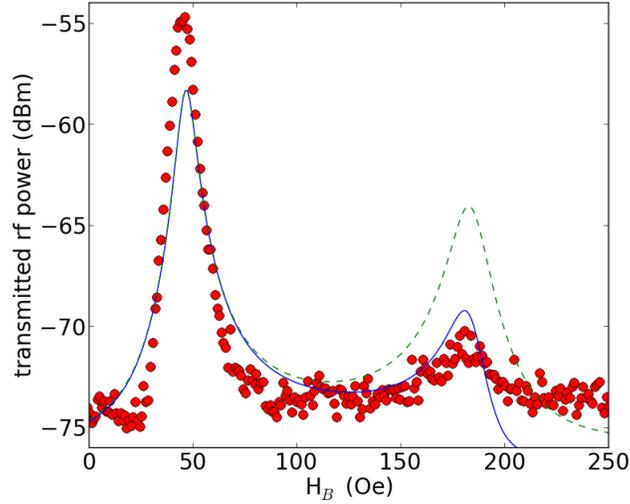}
	\caption{\label{Fig.3} 4 GHz generation with input signal at 2 GHz, +18.9 dBm. A second peak at the bias field for 4 GHz FMR is clearly present; red dots: experimental data; dashed green: theoretical; blue: adjusted theoretical with input rf impurity. See text for details.}
\end{figure}

\textit{Summary}: We have demonstrated a highly efficient frequency doubling effect in thin-film Ni$_{81}$Fe$_{19}$ for input powers well below the Suhl instability threshold. An analysis of the intrinsically nonlinear LLG equation interprets the observed phenomena quantitatively. The results explore new opportunities in the field of rf signal manipulation with CMOS compatible thin film structures. \\

We acknowledge Stephane Auffret for the Ni$_{81}$Fe$_{19}$ sample. We acknowledge support from the US Department of Energy grant DE-EE0002892 and National Science Foundation ECCS-0925829.


\begin{thebibliography}{99}

\bibitem{AyresJAP1956}
W. P. Ayres, P. H. Vartanian, and J. L. Melchor, J. Appl. Phys. \textbf{27}, 188 (1956)

\bibitem{WangPR1954}
N. Bloembergen and S. Wang, Phys. Rev. \textbf{93}, 72 (1954)

\bibitem{SuhlJPCS1957}
H. Suhl, J. Phys. Chem. Solids. \textbf{1}, 209 (1957)

\bibitem{BierleinPRB1970}
J. D. Bierlein and P. M. Richards, Phys. Rev. B \textbf{1}, 4342 (1970)

\bibitem{RidrigueJAP1969}
G. P. Ridrigue, J. Appl. Phys. \textbf{40}, 929 (1969)

\bibitem{HarrisIEEE2012}
V. G. Harris, IEEE Trans. Magn. \textbf{48}, 1075 (2012)

\bibitem{DemidovNMat2012}
V. E. Demidov, S. Urazhdin, H. Ulrichs, V. Tiberkevich, A. Slavin, D. Baither, G. Schmitz and S. O. Demokritov, Nature Mat. \textbf{11}, 1028 (2012)

\bibitem{BerteaudJAP1966}
A. Berteaud and H. Pascard, J. Appl. Phys. \textbf{37}, 2035 (1966)

\bibitem{GerritsPRL2007}
T. Gerrits, P. Krivosik, M. L. Schneider, C. E. Patton, and T. J. Silva, Phys. Rev. Lett. \textbf{98}, 207602 (2007)

\bibitem{OlsonJAP2007}
H. M. Olson, P. Krivosik, K. Srinivasan, and C. E. Patton, J. Appl. Phys. \textbf{102}, 023904 (2007)

\bibitem{BaoAPL2008}
M. Bao, A. Khitun, Y. Wu, J. Lee, K. L. Wang, and A. P. Jacob, Appl. Phys. Lett. \textbf{93}, 072509 (2008)

\bibitem{KhivintsevAPL2011}
Y. Khivintsev, J. Marsh, V. Zagorodnii, I. Harward, J. Lovejoy, P. Krivosik, R. E. Camley, and Z. Celinski, Appl. Phys. Lett. \textbf{98}, 042505 (2011)

\bibitem{MarshAPL2012}
J. Marsh, V. Zagorodnii, Z. Celinski, and R. E. Camley, Appl. Phys. Lett. \textbf{100}, 102404 (2012)

\bibitem{YanaJMMM2008}
M. Yana, P. Vavassori, G. Leaf, F.Y. Fradin, and M. Grimsditch, J. Magn. Magn. Mater \textbf{320}, 1909 (2008)

\bibitem{DemidovAPL2011}
V. E. Demidov, H. Ulrichs, S. Urazhdin, S. O. Demokritov, V. Bessonov, R. Gieniusz, and A. Maziewski, Appl. Phys. Lett. \textbf{99}, 012505 (2011)

\bibitem{BiAPL2011}
C. Bi, X. Fan, L. Pan, X. Kou, J. Wu, Q. Yang, H. Zhang, and J. Q. Xiao, Appl. Phys. Lett. \textbf{99}, 232506 (2011)

\bibitem{BushnellRSI1992}
S. E. Bushnell, W. B. Nowak, S. A. Oliver, and C. Vittoria, Rev. Sci. Instrum. \textbf{63}, 2021 (1992)

\bibitem{Gurevich&Melkov}
A. G. Gurevich and G. A. Melkov, Magnetization Oscillation and Waves (CRC, Boca Raton, 1996)


\end{thebibliography}
\end{document}